\documentstyle[preprint,aps,graphics,graphicx,epsfig]{revtex}
\tightenlines

\newcommand{\ket}[1]{|{#1}\rangle}
\newcommand{\bra}[1]{\langle{#1}|}

\begin{document}

\title{Components for Optical Qubits in the Radio
Frequency Basis} \author{E.H.Huntington} \address{Centre for Quantum
Computer Technology, \\
School of Information Technology and Electrical Engineering, University College,\\
The University
of New South Wales, \\
Canberra ACT 2600 Australia\\
E-mail:e.huntington@adfa.edu.au}
\author{T.C.Ralph}
\address{Centre for Quantum Computer Technology,
\\ Department of Physics, The University of Queensland,
\\ St Lucia QLD 4072 Australia}

\maketitle

\begin{center}
{\scriptsize (3 November 2003)}
\end{center}

\begin{abstract}
    We describe a scheme for the encoding and manipulation of single photon
    qubits in radio frequency sideband modes using standard optical elements.

\end{abstract}

\vspace{10 mm}

%\newpage
\section{Introduction}
Quantum information can be encoded and manipulated using single photon
states. Many in principle demonstrations of quantum information tasks
have now been accomplished in single photon optics including quantum 
cryptography
\cite{butt98}, quantum dense coding \cite{mattle} and quantum teleportation
\cite{bou}. More recently two qubit gates have been realised
\cite{pittman,job}
using conditional techniques \cite{knill}.
Such experiments typically make use of
polarisation to encode the qubits.
However, polarisation is
not the only photonic degree of freedom available to the
experimentalist. For example schemes in which the timing \cite{zbi98} or
occupation \cite{lom} of optical modes are the quantum variables
have also been realized.

We consider here another possibility: a scheme whereby an
optical qubit is encoded in the occupation by a single photon of one
of two different frequency modes.  Two optical frequency basis states,
separated by radio or microwave frequencies, would be sufficiently
close together that they could be manipulated with standard
electro-optical devices but still clearly resolvable using narrowband
optical and opto-electronic systems.  There is also the tantalising
possibility of implementing such a scheme using commercial optical
fibre technologies.  Hence, such an encoding scheme is
attractive from the perspective of developing stable, robust and
ultimately commercially viable optical quantum information systems.

The experimental attraction of developing optical fibre based quantum
optical systems is clear.  For example, there is ongoing interest in
developing non-classical optical sources that will be well suited to
optical fibres and optical fibre technologies
\cite{fiorentino,sharping,silberhorn}.  Indeed a Quantum Key
Distribution (QKD) scheme using radio frequency amplitude and phase
modulation as the conjugate bases has recently been demonstrated using
optical fibres and fibre technologies \cite{merolla}.

If general experiments in the ``radio frequency basis'' (RF basis) are to
become viable, we would require analogies of the tools of the
trade used in polarisation encoding schemes.  These tools are
principally the half-wave plate (HWP) which is used to make arbitrary
rotations of a state in the polarisation basis, the polarising
beamsplitter (PBS) which is used to separate (or combine) photons into
(from) different spatial and polarisation modes and the quarter-wave
plate (QWP) which is used to introduce relative phase shifts
between the two bases.

The principle contribution of this work is to introduce and then
analyse a device which produces arbitrary rotations in the RF basis -
a ``Radio Frequency Half-Wave Plate'' (RF-HWP).  A necessary component
of the RF-HWP is a ``Frequency Beamsplitter'' (FBS) - the RF-basis
analogue of the PBS.  Previous papers have described a device
based on a Fabry-Perot cavity which could be used as a FBS
\cite{hunt02,zhang03}.  Here we shall outline a technique that is far less
experimentally challenging than that discussed in Ref.  \cite{hunt02}.
We note that relative phase shifts (that is, QWP action) can be
achieved through propagation.

\section{In Principle}

In order to illustrate the physics of the device let us first take an idealized
example of an encoding scheme which makes use of the radio frequency
basis.  Consider the logical basis whereby

   \begin{eqnarray}
       \label{basis} \ket{0}_{L}=\ket{1}_{-\Omega} \ket{0}_{+\Omega}
       \nonumber \\
       \ket{1}_{L}=\ket{0}_{-\Omega} \ket{1}_{+\Omega}
       \end{eqnarray}

       \noindent where $\ket{0}_{L}$ and $\ket{1}_{L}$ denote the
       logical states of the qubit and the notation $\ket{n}_{\pm
       \Omega}$ denotes an $n$ photon state at the frequency $\pm
       \Omega$ relative to the average or carrier frequency
       $\omega_{0}$.  We shall assume that the states are
       indistinguishable apart from their frequencies and we note that
       $\Omega$ is taken to be a radio or microwave frequency in the
       range of tens of megahertz to a few gigahertz.  It is convenient
       to write the states of Eq.\ref{basis} in the form

   \begin{eqnarray}
       \label{basis2} \ket{0}_{L}= A(-\Omega)^{\dagger}\ket{0}
       \nonumber \\
       \ket{1}_{L}= A(+\Omega)^{\dagger}\ket{0}
       \end{eqnarray}

\noindent where $A(\omega)^{\dagger}$ is the creation operator for the
$\omega$ frequency mode. As all the elements in our device are
passive (energy conserving) we can obtain the state evolution
produced through the device by considering the Heisenberg evolution of
the relevant operators.

\begin{center} \begin{figure}[htbp] \vspace{0.5cm}
\centerline{ \psfig{figure=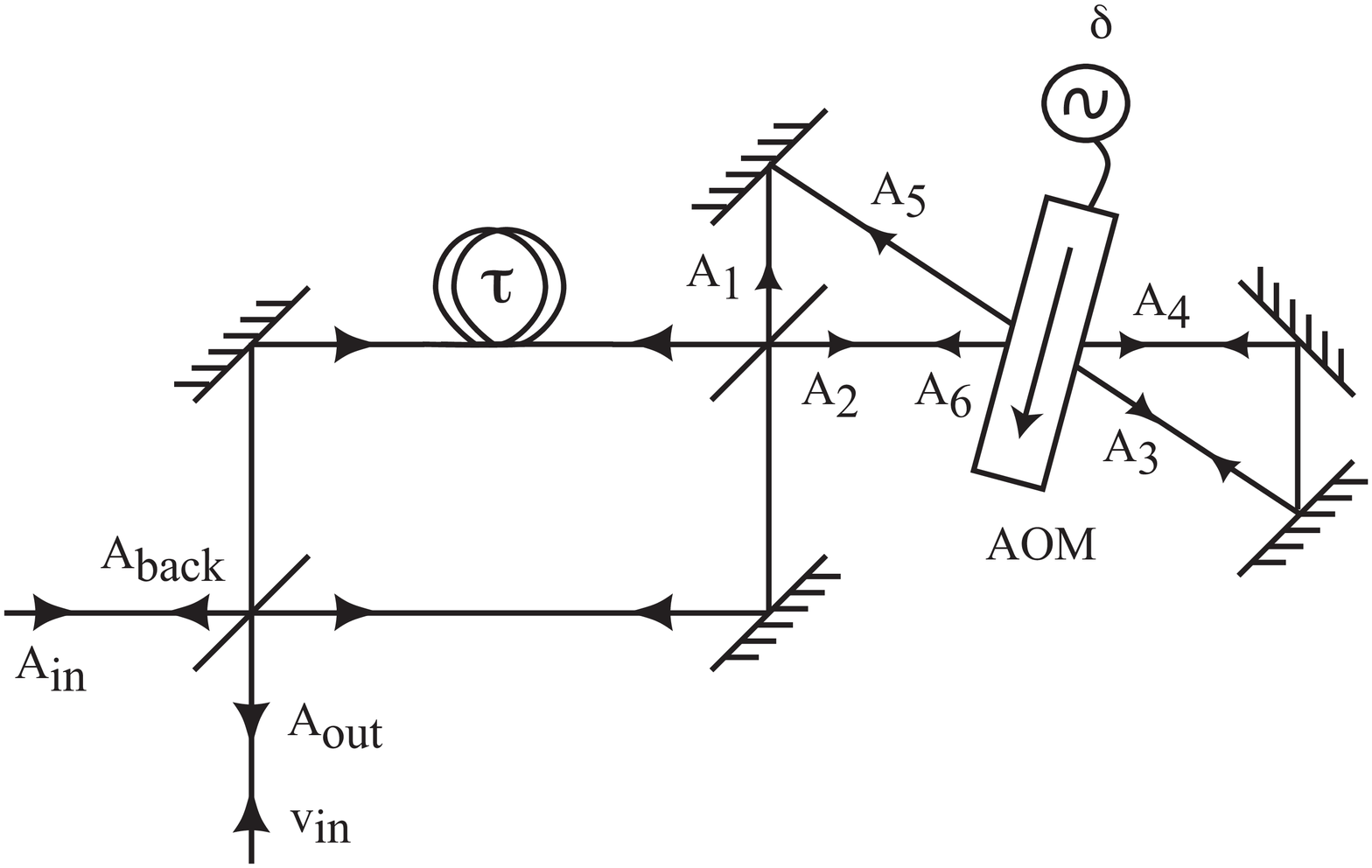,width=15cm} } \vspace{1cm} \caption{
\label{schem} A schematic diagram of the proposed system.  All of the
beamsplitters are $50\%$ transmitting, the internal fields are
labelled as $A_{k}$ and the arrows indicate directions of propagation
of optical fields.  The differential time delay in the two arms of the
Mach-Zehnder interferometer is indicated by the time delay $\tau$.
The abbreviation AOM stands for acousto-optic modulator and the arrow
in the AOM indicates the direction of propagation of the acoustic wave
in the device.  The frequency of the RF source used to
drive the AOM is $\delta$.}
\end{figure}\end{center}

Figure \ref{schem} illustrates schematically the RF-HWP. We shall
assume that all of the beamsplitters are $50\%$ transmitting in our
analysis of the system depicted in Fig.  \ref{schem}.  We shall
additionally make use of the symmetric beamsplitter convention
\cite{walls}.  Let us denote the ``forward travelling'' beams as those
propagating to the right of Fig.  \ref{schem} and the ``backward
travelling'' beams as those propagating to the left.

In the Heisenberg picture, we define the annihilation operator for the
input mode at a particular
Fourier frequency $\omega$ relative to the carrier as
$A_{in}(\omega)$. We also define an ancilla field ${\hat
v}_{in}(t)$, initially in a vacuum state,
entering the device vertically from the bottom.
Fig.  \ref{schem} defines a number of internal annihilation
operators for the RF-HWP as well as two output fields.  The output of
interest is $A_{out}$.

If we take the logical basis as defined in Eq. \ref{basis} and
\ref{basis2}, we become
interested specifically in the operators, $A_{k}(\Omega)$ and
$A_{k}(-\Omega)$ where $k=in,1..6,out,back$.  We shall make use of
the relation that $A_{k}^{\dagger}(\omega)= A_{k}(-\omega)^{\dagger}$
\cite{glauber} to find the relevant creation operators.

The first stage of the RF-HWP is essentially a highly
asymmetric Mach-Zehnder interferometer.  The annihilation operators for the
forward travelling outputs of the Mach-Zehnder interferometer at the
frequency $\omega$ may be written as

\begin{eqnarray}
      \label{FBS1} A_{1}(\omega)&=&\frac{1}{2} \left[ A_{in}(\omega)
      \left(- e^{\imath \phi}e^{\imath \omega \tau} + 1 \right) + \imath
      v_{in}(\omega) \left( e^{\imath \phi} e^{\imath \omega \tau}+1
      \right) \right] \nonumber \\
      \label{FBS2} A_{2}(\omega)&=&\frac{1}{2} \left[ \imath
A_{in}(\omega) \left( e^{\imath
      \phi}e^{\imath \omega \tau} + 1 \right) + v_{in}(\omega) \left(
      e^{\imath \phi} e^{\imath \omega \tau}-1 \right) \right]
      \end{eqnarray}

      \noindent where $A_{1}$ and $A_{2}$ are defined in Fig.
      \ref{schem}, $\tau$ is the differential time delay introduced into
      one of the arms of the interferometer and $\phi=\omega_{0} \tau$ is
      the phase shift acquired by a field at the carrier frequency

Choosing the time delay $\tau$ in the interferometer such that
$\phi=\pi/2$ and $\Omega \tau=\pi/2$, the creation operators for the
forward travelling outputs of the Mach-Zehnder interferometer at the
frequencies $\omega=\pm \Omega$ are

\begin{eqnarray}
      A_{1}(\Omega)^{\dagger}&=&A_{in}(\Omega)^{\dagger}, \quad
      A_{1}(-\Omega)^{\dagger}= -\imath v_{in}(-\Omega)^{\dagger}
      \nonumber \\
      A_{2}(\Omega)^{\dagger}&=&-v_{in}(\Omega)^{\dagger}, \quad
      A_{2}(-\Omega)^{\dagger}=- \imath A_{in}(-\Omega)^{\dagger}
      \nonumber
      \end{eqnarray}

In the state picture we have that an arbitrary input state

\begin{equation}
      \ket{\psi}=(\mu A_{in}(-\Omega)^{\dagger} + \nu
      A_{in}(\Omega)^{\dagger}) \ket{0}_{Ain} \ket{0}_{vin}
      \end{equation}

\noindent is transformed to the output state $\ket{\psi'}$ according
to

\begin{eqnarray}
     \ket{\psi'} & = & U \ket{\psi} = U (\mu
     A_{in}(-\Omega)^{\dagger} + \nu
     A_{in}(\Omega)^{\dagger}) \ket{0}_{Ain} \ket{0}_{vin}
     \nonumber \\
     & = & (\mu U A_{in}(-\Omega)^{\dagger} U^{\dagger} + \nu
     U A_{in}(\Omega)^{\dagger} U^{\dagger}) \ket{0}_{A1}
     \ket{0}_{A2} \nonumber \\
     & = & (\mu  \imath A_{2}(-\Omega)^{\dagger} + \nu
     A_{1}(\Omega)^{\dagger}) \ket{0}_{A1} \ket{0}_{A2}
     \nonumber \\
     & = & \imath \mu \ket{0}_{A1} \ket{1}_{-\Omega,A2} + \nu
     \ket{1}_{+\Omega,A1} \ket{0}_{A2}
\end{eqnarray}
\noindent where $U$ is the unitary operator representing the evolution
through the element. In going from lines two to three we have used
the fact that  $U A_{in}(-\Omega)^{\dagger} U^{\dagger}$ is time
reversed Heisenberg evolution, obtained explicitly by inverting the standard
Heisenberg equations such that the input
operator is written in terms of the output
operators. We have also used $U \ket{0}_{Ain}\ket{0}_{vin} =
\ket{0}_{A1}\ket{0}_{A2}$.  We see that the action of an asymmetric
Mach-Zehnder on the frequency encoding is equivalent to the action of
a polarising beamsplitter in polarisation encoding.

The heart of the RF-HWP is an acousto-optic modulator (AOM).
An AOM couples two different frequency and spatial modes together via
a phonon interaction \cite{resch,young}.  In our scheme the AOM is
used to shift photons between the two frequencies $-\Omega$ and
$\Omega$.  The annihilation operators for the forward travelling
outputs of the AOM, $A_{3}(\omega)$ and $A_{4}(\omega)$ as defined in
Fig.  \ref{schem}, are \cite{resch}

\begin{eqnarray}
      \label{upshift1} A_{3}(\omega)& =&\cos\theta A_{1}(\omega) +
      \imath \sin \theta A_{2}(\omega-\delta) \\
      \label{downshift1} A_{4}(\omega) &=& \cos \theta A_{2}(\omega) +
      \imath \sin \theta A_{1} (\omega + \delta)
      \end{eqnarray}
\noindent where $\delta$ represents the modulation frequency applied
to the AOM and $\theta$ is a measure of the diffraction efficiency of the
AOM such that $\cos \theta$ represents the undiffracted fraction of
the field and $\sin \theta$ represents the diffracted fraction.  We
have taken the asymmetric phase convention for the AOM outputs and
note that $\theta$ is proportional to the amplitude of the radio
frequency modulation applied to the AOM \cite{young}.  We note that
the second term in Eq.  \ref{upshift1} represents the diffracted, and
hence frequency upshifted, component of the input field $A_{2}$.
Similarly, the second term in Eq.  \ref{downshift1} represents the
downshifted component of $A_{1}$.

We double pass the AOM in this scheme.  The two backward travelling
fields $A_{5}$ and $A_{6}$ emerge from the AOM as illustrated in Fig.
\ref{schem}.  The annihilation operators for
the backward travelling fields emerging from the AOM are
\begin{eqnarray}
      \label{downshift2} A_{5}(\omega) &=& \cos \theta A_{4}(\omega) + \imath
      \sin \theta A_{3} (\omega+\delta) \nonumber \\
      &=& \cos 2\theta A_{2}(\omega) + \imath \sin 2\theta
      A_{1}(\omega + \delta) \\
      \label{upshift2} A_{6} (\omega) &=& \cos \theta A_{3} (\omega) +
      \imath \sin \theta A_{4} (\omega - \delta) \nonumber \\
      &=& \cos 2 \theta A_{1}(\omega) + \imath \sin 2 \theta A_{2}
      (\omega - \delta)
\end{eqnarray}

The backward travelling fields $A_{5}$ and $A_{6}$ make a second pass
of the Mach-Zehnder interferometer.  The annihilation operators for
the output fields of the system, $A_{out}$ and $A_{back}$, are

\begin{eqnarray}
      \label{output1inter} A_{out}(\omega)&=&\frac{1}{2} \left[ \imath
      A_{5}(\omega) \left( e^{\imath \phi}e^{\imath \omega \tau} + 1
      \right) + A_{6}(\omega) \left( e^{\imath \phi} e^{\imath \omega
      \tau}-1 \right) \right] \\
      \label{output2inter} A_{back}(\omega)&=&\frac{1}{2} \left[
      A_{5}(\omega) \left(- e^{\imath \phi}e^{\imath \omega \tau} + 1
      \right) + \imath A_{6}(\omega) \left( e^{\imath \phi} e^{\imath
      \omega \tau}+1 \right) \right]
      \end{eqnarray}

\noindent where the parameters $\phi$ and $\tau$ are the same as
those for the forward travelling asymmetric Mach-Zehnder
interferometer.

We can combine the results of Equations \ref{FBS1},
\ref{downshift2}, \ref{upshift2}, \ref{output1inter} and
\ref{output2inter} as well as making use of $A^{\dagger}(\omega) =
A(-\omega)^{\dagger}$ to arrive at the creation operators for the
outputs of the RF-HWP in terms of the inputs.  Focussing on the
downward travelling output $A_{out}$ at the frequencies of interest
$\omega=\pm\Omega$, setting the AOM modulation frequency $\delta = 2
\Omega$ and setting $\tau$ such that $\phi=\pi/2$ and $\Omega
\tau=\pi/2$, we find that
\begin{eqnarray}
      \label{output1b}  A_{out}(-\Omega)^{\dagger}&=& - \left[
      \cos 2 \theta A_{in}(-\Omega)^{\dagger} + \sin 2 \theta
      A_{in}(\Omega)^{\dagger} \right] \nonumber \\
        A_{out}(\Omega)^{\dagger}&=& - \left[
      \cos 2 \theta A_{in}(\Omega)^{\dagger} - \sin 2 \theta
      A_{in}(-\Omega)^{\dagger} \right]
      \end{eqnarray}

Applying Eq.  \ref{output1b} to the $\ket{0}_{L}$
input state leads to

\begin{eqnarray}
      \label{outputs 0}
      \ket{\psi}_{0,out} & = & -\left[
      \cos 2 \theta A_{out}(-\Omega)^{\dagger} - \sin 2 \theta
      A_{out}(\Omega)^{\dagger} \right] \ket{0}_{Aout} \ket{0}_{vout}
      \nonumber \\
      & = & -(\cos 2 \theta \ket{0}_{L} - \sin 2 \theta
      \ket{1}_{L})
      \end{eqnarray}

\noindent whilst applying it to the $\ket{1}_{L}$ input gives

\begin{eqnarray}
      \label{outputs 1}
      \ket{\psi}_{1,out} & = & -\left[
      \cos 2 \theta A_{out}(\Omega)^{\dagger} + \sin 2 \theta
      A_{out}(-\Omega)^{\dagger} \right] \ket{0}_{Aout} \ket{0}_{vout}
      \nonumber \\
      & = & -(\cos 2 \theta \ket{1}_{L} + \sin 2 \theta
      \ket{0}_{L})
      \end{eqnarray}

\noindent Equations \ref{outputs 0} and \ref{outputs 1} are the key
results of this paper.  Up to a global phase, these equations are
formally equivalent to those used to describe the rotation of an
arbitrary two dimensional vector through the angle $\Theta=2\theta$.
Hence the system illustrated in Fig.  \ref{schem} is operationally
equivalent\footnote{A waveplate reflects the polarisation of an
incoming beam about the optic axis rather than performing an arbitrary
rotation through an angle.  Hence technically the device proposed here is more
akin to a Babinet compensator than a waveplate.} to a half-wave
plate in the basis defined by the frequencies $-\Omega$ and $\Omega$.

\section{In Practice}

The situation considered so far is of course impractical, as single
frequency qubits will be stationary in time. More realistically, we
might consider finite bandwidth qubits of the form:

\begin{eqnarray}
       \label{basis3} \ket{0}_{L}& = & ({{2}\over{\pi 
\sigma}})^{1/4}\int d\omega
       e^{-(\Omega +\omega)^{2}/\sigma} A_{in}(\omega)^{\dagger}\ket{0}
       \nonumber \\
       \ket{1}_{L}& = & ({{2}\over{\pi \sigma}})^{1/4} \int d\omega
       e^{-(\Omega -\omega)^{2}/\sigma} A_{in}(\omega)^{\dagger}\ket{0}
       \end{eqnarray}

\noindent The overlap between these qubits is

\begin{equation}
      \langle 0 | 1 \rangle_{L} = e^{-2 \Omega^{2}/\sigma}
      \end{equation}

\noindent thus provided the width of the frequency packet is
sufficiently small compared to its mean (i.e $\sigma << 2 \Omega^{2}$)
then these qubits will be approximately orthogonal.  The problem with
the finite frequency spread for our device is that now the condition
$\Omega \tau=\pi/2$ will not be precisely satified for the entire
frequency packet.  The effect is to produce a phase shift across the
frequency wave packet and also to produce some probability of photons
exiting the device in the wrong beam (ie $A_{back}$) or at frequencies
outside the computational basis (eg $2 \Omega$).  Taking such events
as lost photons and tracing over them leads to a mixed state which can
be written

\begin{equation}
      \rho_{out,i} = \ket{Q}_{i}\bra{Q}_{i} + \ket{\bar Q}\bra{\bar Q}
      \end{equation}

\noindent where $\rho_{out,1}$ represents the mixed state output
obtained for the logical state input $\ket{i}_{L}$.  For the
$\ket{0}_{L}$ input state

\begin{eqnarray}
      \ket{Q}_{0} & = & -({{2}\over{\pi \sigma}})^{1/4}
      \int d\omega e^{-(\Omega+\omega)^{2}/\sigma}(\cos 2
      \theta A_{in}(\omega)^{\dagger} 1/2(1+e^{i
      \pi(\Omega+\omega)/\Omega})- \nonumber\\
    & &  \sin 2
      \theta A_{in}(\omega+2 \Omega)^{\dagger} 1/4(1+e^{i
      \pi/2(\Omega+\omega)/\Omega})^{2})\ket{0}
      \end{eqnarray}

\noindent and for the $\ket{1}_{L}$ input

\begin{eqnarray}
      \ket{Q}_{1} & = & -({{2}\over{\pi \sigma}})^{1/4}
      \int d\omega e^{-(\Omega-\omega)^{2}/\sigma}(\cos 2
      \theta A_{in}(\omega)^{\dagger} 1/2(1+e^{-i
      \pi(\Omega-\omega)/\Omega})+ \nonumber\\
   & &   \sin 2
      \theta A_{in}(\omega-2 \Omega)^{\dagger} 1/4(1+e^{-i
      \pi/2(\Omega-\omega)/\Omega})^{2})\ket{0}
      \end{eqnarray}

\noindent and where $\ket{\bar Q}$ is a collective ket representing
all the photons that end up in (orthogonal) states outside the
computational basis. We can evaluate the impact of this effect by
calculating the fidelity between the expected output state,
$\ket{P}_{i}$,
and that obtained:

\begin{equation}
      F=\bra{P}\rho_{out,i} \ket{P}_{i}
      \end{equation}

\noindent The expression for the fidelity is complicated and of limited
utility to reproduce here, however its pertinent features can be
listed sucinctly: the fidelity depends strongly on the ratio of
$\Omega^{2}$ to $\sigma$; it depends weakly on the rotation angle $\theta$;
as expected it tends to one as the ratio $\Omega^{2}/\sigma$ tends to
infinity. Some representative results are: $\Omega^{2}/\sigma = 10$,
$F = 0.942$; $\Omega^{2}/\sigma = 100$, $F = 0.9939$;
$\Omega^{2}/\sigma = 1000$, $F = 0.99938$. We conclude that high
fidelities are consistent with sensible signal bandwidths.

Let us now turn to technical issues.  Conceptually, the RF-HWP
comprises a FBS, followed by an AOM followed
by another FBS. However, implementing the RF-HWP in such a fashion
would be a tremendously challenging technical task.  It would require
actively locking the phase $\phi=\pi/2$ in two different
interferometers.  In addition, the optical path length between the two
interferometers would need to be locked.  This is why we have proposed the
RF-HWP with the folded design illustrated in Fig.  \ref{schem}.
This folded design requires locking of only one interferometer.
Further, a locking signal can be derived from the backward travelling
output of the RF-HWP, $A_{back}$, without disturbing the useful
output $A_{out}$.

The technical limitations to the performance of the RF-HWP will be set
by the diffraction efficiency of the AOM, the transmission losses of
the AOM and the mode-matching efficiency in the interferometer.
Rotation of the input through an angle of $\Theta=\pi/2$ requires that
the AOM have a diffraction efficiency of $50 \%$.  This technical
requirement can be met with commercially available devices
\cite{brimrose}.  Transmission losses in the AOM may be modelled as a
perfectly transmitting AOM with a partially transmitting beamsplitter
placed on each output port.  We model mode-matching efficiency in the
interferometer similarly.  Using this approach, and noting that both
the AOM and interferometer are double-passed, we find that the output
of the RF-HWP may be written as \begin{equation}
\rho_{out}^{\prime}=\eta_{T} \rho_{out} + (1-\eta_{T}) \ket{0}\bra{0}
\end{equation}

\noindent where $\rho_{out}^{\prime}$ represents the new output of the
RF-HWP after taking account of transmission loss in the AOM
($\eta_{AOM}$) and the mode-matching efficiency of the interferometer
($\eta_{mm}$) \cite{lvovsky}.  The total transmission of the RF-HWP is
$\eta_{T}=\eta_{AOM}^{2}\eta_{mm}^{2}$.  We make use of a collective
ket $\ket{0}$ to represent the vacua introduced within the device.  A
good quality free-space AOM would have $\eta_{AOM} > 0.95$
\cite{brimrose}.  Similarly, a well mode-matched interferometer would
have $\eta_{mm} > 0.95$ \cite{buc01}.  It would therefore be realistic
to expect that the RF-HWP could attain total transmission of $\eta_{T}
\approx 0.81$ with current commercially available technologies.  The
effective single-pass transmission of the FBS proposed here would be
$\eta_{mm}$.

\section{Conclusion}

In summary, we have proposed devices which may be used as the
principle experimental components in optical quantum
imformation systems which make use of the radio frequency basis.
These components are essentially the RF basis analogues of the
polarising beamsplitter and the half-wave plate.  We have shown that
an asymmetric Mach-Zehnder interferometer can perform the function of
a frequency beamsplitter.  We have also shown that this system may
be combined with an acousto-optic modulator in a folded design
to form a radio frequency half-wave plate.  We have shown
both devices are feasible using current technologies and could operate
with reasonable bandwidths.

\acknowledgments We wish to acknowledge many fruitful discussions with
G. N. Milford.  This work was supported by the Australian Research
Council.  

\newpage
\begin{thebibliography}{99}

\bibitem{butt98} W.~T.~Buttler et al, \pra {\bf 57}, 2379 (1998).

\bibitem{mattle} K.~Mattle, H.~Weinfurter, P.~G.~Kwiat, and A.~Zeilinger
\prl {\bf 76}, 4656-4659 (1996).

\bibitem{bou} D.~Bouwmeester,
  J.~W.~Pan, K.~Mattle,
  		M.~Eibl, H.~Weinfurter and A.~Zeilinger,
  Nature {\bf 390}, 575 (1997).

\bibitem{pittman} T. B. Pittman, B. C.
Jacobs, J. D. Franson, \prl, {\bf 88}, 257902 (2002);
T.~B.~Pittman,  M.~J.~Fitch, B.~C.~Jacobs and J.~D.~Franson,
quant-ph/0303113 (2003).

\bibitem{job} J.L.O'Brien, G.J.Pryde, A.G.White, T.C.Ralph and
D.Branning, submitted (2003).

      %klm
\bibitem{knill}
E. Knill, R. Laflamme, and G. J. Milburn, Nature, {\bf 404}, 48
(2001).


\bibitem{zbi98} H.~Zbinden et al, Appl.Phys.B {\bf 67}, 743 (1998).

\bibitem{lom} E.~Lombardi, F.~Sciarrino, S.~Popescu, and F.~De~Martini
\prl {\bf 88}, 070402 (2002)


%kumar fibre sources
\bibitem{fiorentino}
M. Fiorentino, P. L. Voss, J. E. Sharping, and P. Kumar, IEEE Ph.
Technol. Lett., {\bf 14}, 983 (2002).

%another kumar fibre source

\bibitem{sharping}
J. E. Sharping, M. Fiorentino, P. Kumar, Opt.  Lett., {\bf 26}, 367
(2001)

%erlangen scheme

\bibitem{silberhorn}
C. Silberhorn, P. K. Lam, O. Weiss, F. Konig, N. Korolkova, G. Leuchs,
\prl, {\bf 86}, 4267 (2001).


%rf qkd
\bibitem{merolla}
J.-M. Merolla, L. Duraffourg, J.-P. Goedgebuer, A. Soujaeff, F.
Patois, and W. T. Rhodes, Eur. Phys. J. D, {\bf 18}, 141 (2002).

      % % cavity paper
\bibitem{hunt02}
%''Separating the quantum sidebands of an
%optical field''
E. H. Huntington and T. C. Ralph
J. Opt. B {\bf 4}, 123 (2002).
% newer cavity paper
\bibitem{zhang03} J. Zhang Phys.  Rev.  A, {\bf 67}, 054302 (2003)
%walls and milburn

\bibitem{walls}
D. F. Walls, and G. J. Milburn, {\it Quantum Optics}, Springer,
Berlin (1995).

%glauber - reference for frequency dependent annihilation operator
\bibitem{glauber}
R. J. Glauber, Phys. Rev., {\bf 130}, 2529 (1963).

%aom comment
\bibitem{resch}
% %``Comment on ``Manipulating the frequency-entangled states by an
%acoustic-optical modulator'',
   K.J.Resch, S. H. Myrskog, J. S. Lundeen, and A. M. Steinberg,
\pra, {\bf 64}, 056101 (2001).
%yariv
\bibitem{young}
E. H. Young, S.-K. Yao, Proc. IEEE, {\bf 69}, 54 (1981).

%lvovsky citation for efficiency
\bibitem{lvovsky}
A. I. Lvovsky, H. Hansen, T. Aichele, O. Benson, J. Mlynek and S.
Schiller, \prl, {\bf 87}, 050402 (2001).

%brimrose citation for AOM
\bibitem{brimrose}
See for example http://www.brimrose.com


%ben
\bibitem{buc01} B.~C.~Buchler, P.~K.~Lam, Hans-A.~Bachor, U.~L.~Andersen,
   and T.~C.~Ralph, \pra {\bf 65}, 011803 (2002)

\end{thebibliography}
\end{document}